\documentclass{aastex701}

\begin{document}

\title{Modeling (Sub-)millimeter Scattering Properties of Fractal and Consolidated Porous Particles: Applications to Protoplanetary Disks}

\correspondingauthor{Gonzalo Vargas Pel\'aez}

\author[0009-0000-5520-4455]{Gonzalo Vargas}
\affiliation{Instituto de Astrof\'{\i}sica de Andaluc\'{\i}a (IAA-CSIC), Glorieta de la Astronom\'{\i}a s/n, E-18008 Granada, Spain}
\email[show]{gvargas@iaa.es}

\author[0000-0002-9228-1035]{Daniel Guirado}
\affiliation{Instituto de Astrof\'{\i}sica de Andaluc\'{\i}a (IAA-CSIC), Glorieta de la Astronom\'{\i}a s/n, E-18008 Granada, Spain}
\email{dani@iaa.es}

\author[0000-0003-2862-5363]{Carlos Carrasco-Gonz\'alez}
\affiliation{Instituto de Radioastronom\'{\i}a y Astrof\'{\i}sica, Universidad Nacional Aut\'onoma de M\'exico, Antigua Carretera a P\'atzcuaro \# 8701, Ex-Hda. San Jos\'e de la Huerta, Morelia, Michoac\'an, M\'exico C.P. 58089}
\email{c.carrasco@irya.unam.mx}

\author[0000-0002-5138-3932]{Olga Mu\~noz}
\affiliation{Instituto de Astrof\'{\i}sica de Andaluc\'{\i}a (IAA-CSIC), Glorieta de la Astronom\'{\i}a s/n, E-18008 Granada, Spain}
\email{olga@iaa.es}

\author[0000-0002-3524-0093]{Maxim A. Yurkin}
\affiliation{Universit\'e Rouen Normandie, INSA Rouen Normandie, CNRS, CORIA UMR 6614, Campus du Madrillet, 675 avenue de l'Universit\'e, BP12, 76801 Saint-\'Etienne-du-Rouvray cedex, France}
\email{yurkin@gmail.com}

\author[0000-0003-1283-6262]{Enrique Mac\'{\i}as}
\affiliation{European Southern Observatory (ESO), Karl-Schwarzschild-Str.\ 2, 85748 Garching bei M\"unchen, Germany}
\email{enrique.macias@eso.org}

\author[0000-0002-1912-5394]{Jes\'us M.\ J\'aquez-Dom\'{\i}nguez}
\affiliation{Instituto de Radioastronom\'{\i}a y Astrof\'{\i}sica, Universidad Nacional Aut\'onoma de M\'exico, Antigua Carretera a P\'atzcuaro \# 8701, Ex-Hda. San Jos\'e de la Huerta, Morelia, Michoac\'an, M\'exico C.P. 58089}
\email{j.jaquez@irya.unam.mx}

\author[0000-0002-0246-4648]{Francisco J.\ Garc\'{\i}a-Izquierdo}
\affiliation{Instituto de Astrof\'{\i}sica de Andaluc\'{\i}a (IAA-CSIC), Glorieta de la Astronom\'{\i}a s/n, E-18008 Granada, Spain}
\email{fgarcia@iaa.es}

\begin{abstract}
The early evolution of dust in protoplanetary disks remains uncertain, as the maximum particle size inferred from (sub-)millimeter polarization can differ by up to an order of magnitude from that inferred from spectral energy distribution modeling. To test whether porosity and morphology can reduce this tension, we perform light-scattering numerical simulations for two dust populations: (i) consolidated porous particles computed with the discrete dipole approximation (\texttt{ADDA}) and (ii) highly porous aggregate models, including fractal and hierarchical aggregates, computed with the multiple-sphere $T$-matrix method (\texttt{MSTM}). Using DSHARP optical constants, we compute scattering matrices, cross sections, and effective albedo $\bar{\omega}_{\mathrm{eff}}$ for a size distribution $n(a)\propto a^q$ with $q=-3.5$, $a_{\min}=0.1\,\mu\mathrm{m}$, and ten wavelengths from $0.87$ to $10\,\mathrm{mm}$. We find that increasing porosity strengthens forward scattering and enhances polarization near $\theta\approx 90^\circ$. For compact spheres, $P(90^\circ)\,\bar{\omega}_{\mathrm{eff}}$ peaks near $a_{\max}\sim \lambda/2\pi$ and then declines, whereas porous particles show a broader peak extending to larger sizes, keeping polarization-based constraints compatible with $a_{\max}\sim 1\,\mathrm{mm}$. Porosity also lowers $\kappa_{\mathrm{abs}}$ at fixed dust mass relative to compact spheres, implying larger inferred dust masses for a given continuum flux.
\end{abstract}

\keywords{\uat{Optical constants (Dust)}{2270} --- \uat{Polarimetry}{1278} --- \uat{Protoplanetary disks}{1300}}

\section{Introduction}\label{sec:introduction}

Protoplanetary disks are gas-rich structures around young stars. Dust sets the (sub-)millimeter continuum opacity, shapes continuum emission and polarization, and provides the building blocks of planet formation \citep{Pinilla2012_trapping,Dullemond2018_DSHARP_VI,Birnstiel2018,NajitaKenyon2014_MNRAS445_3315}. High-angular-resolution observations, particularly with the Atacama Large Millimeter/submillimeter Array (ALMA) and the NSF's Karl G.\ Jansky Very Large Array (VLA), reveal rings, gaps, and azimuthal asymmetries that can trap solids and promote particle growth (see \citealt{Andrews2020}).

At (sub-)millimeter wavelengths, scattering can contribute significantly to continuum emission and polarization. Neglecting scattering can bias inferred extinction opacities and lead to substantial underestimates of disk masses, especially in optically thick regions and when characteristic particle sizes are comparable to the observing wavelength \citep{Carrasco2019,Sierra2020ApJ}. The interpretation of these observables is further complicated by degeneracies among particle size, morphology, composition, and disk structure. HL~Tau illustrates this difficulty: polarization measurements have been interpreted to favor $a_{\max}\sim 160\,\mu\mathrm{m}$ \citep{Kataoka2015}, whereas the continuum spectrum has often been modeled with $a_{\max}\sim 1\,\mathrm{mm}$ compact spherical particles \citep[e.g.,][]{Ricci2010,Carrasco2019}.

The assumption that dust particles can be represented as compact homogeneous spheres is computationally convenient, but observational studies increasingly point to dust populations with complex internal structure and irregular morphology \citep[see][]{Ginski2023}, while dust-growth models predict that coagulation naturally produces porous aggregates \citep{Weidenschilling1977,Brauer2008,Testi2014,Okuzumi2012,Suyama2012,Estrada2022,Birnstiel2024ARAA}. Several studies have considered compact nonspherical particles, including spheroidal and irregular shapes \citep[e.g.,][]{KirchschlagerWolf2014,Tazaki2016ApJ823_70,KirchschlagerEtAl2019_intrinsic_porosity,Lin2024}. This suggests that a more substantial departure from the compact-sphere approximation is required, with porosity being a natural candidate.

Accurate light-scattering calculations for irregular porous particles are therefore needed to compare models with observations to constrain $a_{\max}$ and particle morphology. We present a porous-particle modeling framework and quantify how porosity affects light scattering and linear polarization. Section~\ref{sec:Dust_in_proto} introduces the relevant fundamentals and summarizes dust models used in previous studies. Section~\ref{sec:methodology} describes the particle populations and numerical methods. Section~\ref{sec:Results} presents the results and compares them with the compact-sphere (CS) reference model. Finally, we summarize our findings in Section~\ref{sec:Conclusions}.
\section{Dust in Protoplanetary Disks}\label{sec:Dust_in_proto}

\subsection{Light Scattering Fundamentals} \label{subsec:scattering_fundamentals}

The interaction of dust with radiation is described by the scattering matrix formalism. For randomly oriented, mirror-symmetric particles, the Stokes vector of the scattered radiation, $(I,Q,U,V)_{\mathrm{sca}}$, is related to that of the incident radiation, $(I,Q,U,V)_{\mathrm{inc}}$, through the scattering matrix $Z_{ij}(\theta)$, where $\theta$ is the scattering angle. In this case, $Z_{ij}$ has six independent elements \citep[see][for details]{Mishchenko2000}:
\begin{equation}
\left(
\begin{array}{c}
I \\
Q \\
U \\
V
\end{array}
\right)_{\mathrm{sca}}
\propto
\left(
\begin{array}{cccc}
Z_{11} & Z_{12} & 0 & 0 \\
Z_{12} & Z_{22} & 0 & 0 \\
0 & 0 & Z_{33} & Z_{34} \\
0 & 0 & -Z_{34} & Z_{44}
\end{array}
\right)
\left(
\begin{array}{c}
I \\
Q \\
U \\
V
\end{array}
\right)_{\mathrm{inc}}.
\label{eq:SCAT_MAT}
\end{equation}

Using the normalized phase matrix notation $F_{ij}(\theta)$, $F_{11}(\theta)$ is the phase function for unpolarized incident light and satisfies
\begin{equation}
\frac{1}{2}\int_0^{\pi} F_{11}(\theta)\,\sin\theta\,\mathrm{d}\theta = 1,
\end{equation}
while the linear polarization fraction is
\begin{equation}
P(\theta) = -\frac{F_{12}(\theta)}{F_{11}(\theta)}.
\label{eq:LPF}
\end{equation}
The scattering behavior depends mainly on the size parameter $x \equiv 2\pi a/\lambda$, where $a$ is the particle size and $\lambda$ is the wavelength. In the Rayleigh regime ($x \ll 1$), particles have a high $P(\theta)$ that peaks near $\theta = 90^\circ$, and the scattering is weak and nearly isotropic, so particles in this regime are hard to detect. Near the Mie/interference regime ($x \sim 1$), particles with $a \simeq \lambda/2\pi$ scatter more light than small particles and have high $P(\theta)$, making this regime central to (sub-)millimeter disk polarimetry. In the geometric optics regime ($x \gg 1$), scattering becomes increasingly peaked toward forward angles, and $P(\theta)$ is often reduced at intermediate scattering angles; however, this reduction is not a general rule, since highly opaque particles can produce strong polarization through surface reflection when the corresponding reflection angle is close to the Brewster angle \citep[Section~5.4.4]{Hapke2012}. For large compact particles, the scattering behavior also depends on the absorptive part of the refractive index \citep[e.g.,][]{Bohren1983,Hansen1974,Munoz2020ApJS}.

\begin{figure*}[ht!]
\centering
\includegraphics[width=17cm]{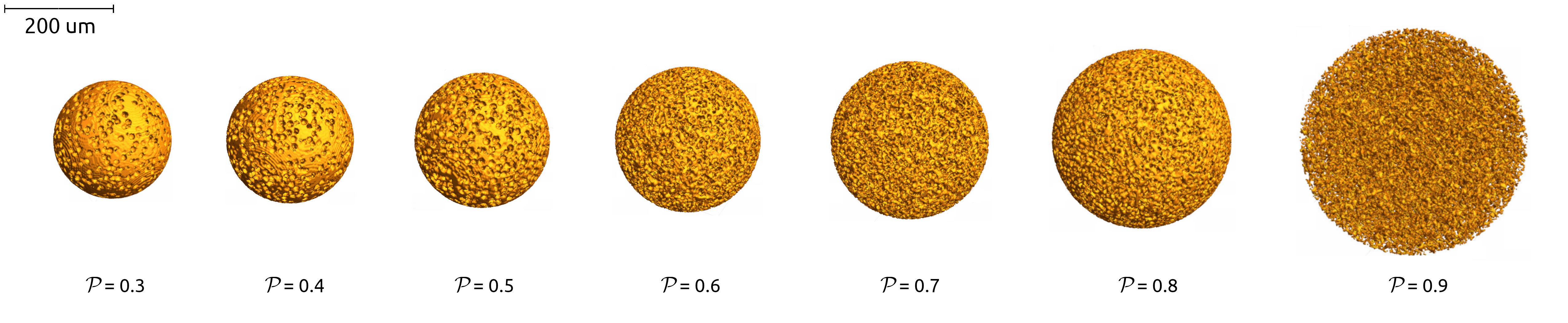}
\caption{Representative consolidated porous (CP) particle models computed with \texttt{ADDA}. Porosity increases from left to right, from $\mathcal{P}=0.3$ to $0.9$. All particles are scaled to the same radius of an equivalent-volume sphere, $a=100\,\mu\mathrm{m}$.}
\label{fig:consolidated_porous}
\end{figure*}

\subsection{Cross Sections and Mass Opacities} \label{subsec:cross_sections}

The scattering, absorption, and extinction cross sections relate the incident irradiance $I_0$ to the corresponding scattered, absorbed, and extinguished powers:
$P_{\mathrm{sca}} = I_0 C_{\mathrm{sca}}$,
$P_{\mathrm{abs}} = I_0 C_{\mathrm{abs}}$, and
$P_{\mathrm{ext}} = I_0 C_{\mathrm{ext}}$.
They satisfy
\begin{equation}
C_{\mathrm{ext}} = C_{\mathrm{sca}} + C_{\mathrm{abs}}.
\label{eq:Cext}
\end{equation}

For unpolarized incidence, $Z_{11}(\theta)$ is the differential scattering cross section, so
\begin{equation}
C_{\mathrm{sca}} = \int_{4 \pi} Z_{11}(\theta)\,\mathrm{d}\Omega
= 2\pi \int_{0}^{\pi} Z_{11}(\theta)\,\sin\theta\,\mathrm{d}\theta.
\label{eq:Csca}
\end{equation}
The corresponding mass opacities are
\begin{equation}
\kappa_{\mathrm{sca}} = \frac{C_{\mathrm{sca}}}{m_{\mathrm{dust}}}, \quad
\kappa_{\mathrm{abs}} = \frac{C_{\mathrm{abs}}}{m_{\mathrm{dust}}}, \quad
\kappa_{\mathrm{ext}} = \frac{C_{\mathrm{ext}}}{m_{\mathrm{dust}}},
\label{eq:Opacities}
\end{equation}

where $m_{\mathrm{dust}}$ is the mass of the particle. The single-scattering albedo is
\begin{equation}
\bar{\omega} = \frac{C_{\mathrm{sca}}}{C_{\mathrm{ext}}}
= \frac{\kappa_{\mathrm{sca}}}{\kappa_{\mathrm{ext}}}.
\label{eq:Albedo}
\end{equation}

For strongly anisotropic scattering, $\bar{\omega}$ can overestimate the contribution of scattering to the emergent intensity because forward-scattered photons only weakly redistribute the radiation field. We quantify this effect with the asymmetry parameter
\begin{equation}
g \equiv \langle \cos\theta\rangle
= \frac{1}{2}\int_{0}^{\pi} F_{11}(\theta)\,\cos\theta\,\sin\theta\,\mathrm{d}\theta.
\label{eq:asymmetry_parameter}
\end{equation}
We then define the effective scattering opacity \citep[e.g.,][]{Birnstiel2018,Tazaki2019}
\begin{equation}
\kappa_{\mathrm{sca,eff}} \equiv \kappa_{\mathrm{sca}}(1-g),
\label{eq:kappa_sca_eff}
\end{equation}
and the effective albedo
\begin{equation}
\bar{\omega}_{\mathrm{eff}} \equiv
\frac{\kappa_{\mathrm{sca,eff}}}{\kappa_{\mathrm{ext,eff}}},
\label{eq:albedo_eff}
\end{equation}
with $\kappa_{\mathrm{ext,eff}}=\kappa_{\mathrm{sca,eff}}+\kappa_{\mathrm{abs}}$. Throughout this work, we use $P(90^\circ)\,\bar{\omega}_{\mathrm{eff}}$ as a proxy for the detectability of polarized scattered emission \citep[e.g.,][]{Tazaki2019,Zhang2023}.

\subsection{Size-averaged Scattering Parameters}
\label{subsec:averaged_parameters}

We adopt a power-law size distribution
\begin{equation}
n(a) \propto a^{q}, \quad a_{\min} \leq a \leq a_{\max},
\label{eq:SD}
\end{equation}
where $n(a)\,\mathrm{d}a$ is the number density of particles with radii in the interval $[a,a+\mathrm{d}a]$. The slope $q=-2.5$ is typical of distributions limited by drift \citep{DAlessio2001}, whereas $q=-3.5$ is characteristic of populations dominated by fragmentation \citep{Mathis1977}, and is the value adopted in this study. We compute size-averaged optical properties by integrating over the size distribution. The size-averaged scattering matrix elements are
\begin{equation}
\langle F_{ij}(\theta) \rangle =
\frac{\int_{a_{\min}}^{a_{\max}} n(a)\, C_{\mathrm{sca}}(a)\, F_{ij}(a,\theta)\, \mathrm{d}a}
{\int_{a_{\min}}^{a_{\max}} n(a)\, C_{\mathrm{sca}}(a)\, \mathrm{d}a},
\label{eq:SD_scatmat}
\end{equation}
The corresponding size-averaged scattering opacity (and analogously $\langle\kappa_{\mathrm{abs}}\rangle$ and $\langle\kappa_{\mathrm{ext}}\rangle$) is
\begin{equation}
\langle \kappa_{\mathrm{sca}} \rangle
= \frac{\int_{a_{\min}}^{a_{\max}} n(a)\, C_{\mathrm{sca}}(a)\,\mathrm{d}a}
{\int_{a_{\min}}^{a_{\max}} n(a)\, m_{\mathrm{dust}}(a)\,\mathrm{d}a}.
\label{eq:SD_kappa_sca}
\end{equation}
The size-averaged extinction satisfies $\langle \kappa_{\mathrm{ext}} \rangle = \langle \kappa_{\mathrm{sca}} \rangle + \langle \kappa_{\mathrm{abs}} \rangle$. We compute $g$ from $\langle F_{11}(\theta) \rangle$ and then evaluate $\langle \kappa_{\mathrm{sca,eff}} \rangle$ and $\langle \bar{\omega}_{\mathrm{eff}} \rangle$ with the same definitions used for single sizes (Equations~\ref{eq:kappa_sca_eff} and~\ref{eq:albedo_eff}). For simplicity, angle brackets are omitted below for size-averaged quantities.

\begin{figure*}[ht!]
\centering
\includegraphics[width=15cm]{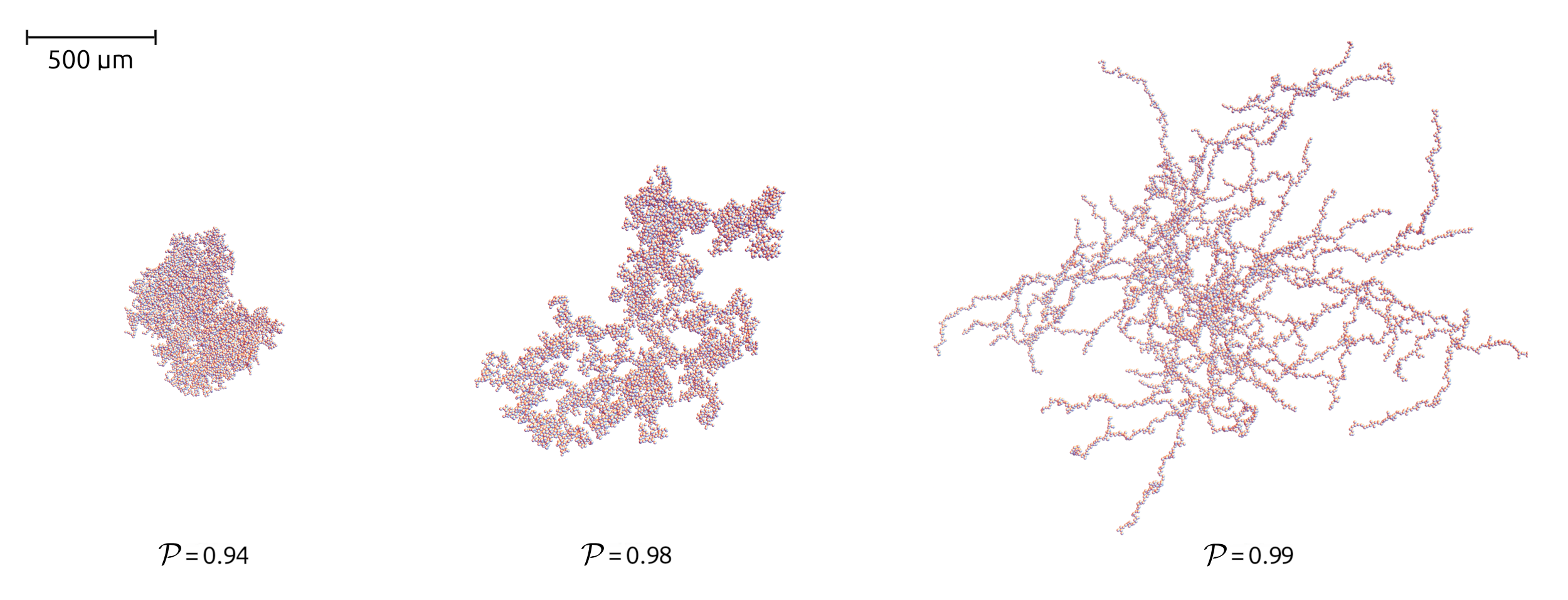}
\caption{Representative aggregate models computed with MSTM. From left to right: Case I, a fractal aggregate (FA) with $D_{\rm f}=2.7$ and $k_{\rm f}=0.3$; Case II, an FA with $D_{\rm f}=2.1$ and $k_{\rm f}=0.7$; and Case III, a hierarchical aggregate (HA) built from 64 randomly oriented sub-aggregates, each containing 128 monomers with $D_{\rm f}=1.2$ and $k_{\rm f}=1.0$. All three cases have the same radius of an equivalent-volume sphere, $a=100\,\mu\mathrm{m}$.}
\label{fig:fractal_aggregates}
\end{figure*}

\subsection{Dust Models in Previous Studies} \label{subsec:dust_models}

Continuum polarization in protoplanetary disks is usually interpreted as a combination of two mechanisms: (i) scattering of thermal dust emission and (ii) dichroic emission from aligned nonspherical particles. At optical and near-infrared wavelengths, polarization is dominated by scattering by small particles \citep{Avenhaus2018}. In the mid-infrared, it is often interpreted as dichroic thermal emission from particles aligned with magnetic or radiative fields \citep{Andersson2015ARAA,Tazaki2017ApJ839_56}. At (sub-)millimeter wavelengths, self-scattering becomes important when particle sizes approach the observing wavelength \citep[e.g.,][]{Kataoka2015,Yang2016a}. HL~Tau illustrates this transition: early detections were interpreted as magnetic alignment \citep{Stephens2014}, whereas later ALMA multi-wavelength data showed morphologies consistent with scattering dominating at shorter wavelengths and alignment contributing at longer wavelengths \citep[e.g.,][]{Kataoka2017ApJL844_L5,Ueda2021,Stephens2023,Lin2023,Lin2024}.

The effective medium approximation (EMA) is widely used to study the effect of porosity through the effective refractive index. EMA-based studies show that high porosity can increase $\kappa_{\mathrm{sca}}$ and $P(\theta)$ while decreasing $\kappa_{\mathrm{abs}}$, with direct consequences for inferred particle sizes and dust masses \citep[e.g.,][]{Liu2022,Liu2024,Zhang2023}. For example, \citet{Zagaria2025} fitted multi-frequency ALMA and VLA data of CI~Tau with EMA particles and favored porosities of $\mathcal{P}\lesssim 0.7$. Another approximation is the distribution of hollow spheres (DHS), which reproduces several porosity signatures, including broadened and shifted spectral features \citep{Min2016}. Particle morphology is an additional degree of freedom \citep[e.g.,][]{KirchschlagerBertrang2020}: DDA calculations for aligned oblate spheroids produce stronger self-scattering polarization than compact spheres near $a\sim \lambda/2\pi$ \citep{KirchschlagerBertrang2020}, and porous prolate/oblate particles can match the HD~142527 polarization for moderate elongations and $\mathcal{P}\lesssim 0.7$ \citep{KirchschlagerEtAl2019_intrinsic_porosity}. More recently, \citet{JaquezDominguez2026} compared radiative-transfer models using solid compact spheres and solid irregular hexahedral particles with identical size distributions, finding that grain irregularity can enhance the scattering opacity and suppress the Mie polarization reversal at large size parameters, although irregularity alone is not sufficient to reproduce the observed millimeter polarization fractions.

Fractal aggregates are a standard description of highly porous particles formed by hit-and-stick growth. \citet{Tazaki2019} showed that, for aggregates composed of submicron monomers, millimeter-wave scattering polarization favors relatively compact structures: aggregates with higher fractal dimension and/or lower porosity are more efficient scatterers, whereas very fluffy aggregates are less likely to contribute significantly to the observed polarized emission. Using optical and near-infrared polarimetry, \citet{TazakiDominik2022} inferred an upper limit of $\sim 0.4\,\mu\mathrm{m}$ for the monomer radius. \citet{Tazaki2023} similarly found that the infrared scattered light of IM~Lup is best reproduced by aggregates with submicron monomers. \citet{Ueda2024_porous_IMLup,Ueda2025ApJ990_183} found that moderately porous dust is favored to reproduce millimeter continuum and polarization simultaneously. These results are consistent with \citet{Ginski2023}, who identified two broad dust populations ranging from highly porous aggregates to moderately porous particles. Independent support comes from CODULAB laboratory scattering measurements \citep{Munoz2010JQSRT}. Based on these data, \citet{Munoz2020ApJS} reproduced the photometric and polarimetric properties of comet 67P/Churyumov--Gerasimenko using millimeter-sized porous clumps composed of submicron particles \citep{Moreno2018}. For a broader synthesis of observational, laboratory, and numerical constraints, see \citet{Potapov2025PorosityReview}.

\begin{table}[ht]
\centering
\caption{Explored parameter space. For the \texttt{MSTM} aggregate models, the adopted maximum particle size depends on $\lambda$.}
\label{tab:param_space}
\begin{tabular}{l c l}
\hline
\multicolumn{1}{c}{Quantity} & \multicolumn{1}{c}{Cases} & \multicolumn{1}{c}{Values} \\
\hline
Wavelength $\lambda$ [$\mathrm{mm}$] & 10 &
$0.87$, $1.0$, $1.25$, $1.6$, $2.0$, $3.0$, $4.0$, $7.0$, $9.0$, $10.0$ \\

Morphology & 2 & \texttt{ADDA} consolidated porous particles (CP) \\
 & & \texttt{MSTM} aggregate models \\
 & & \quad fractal aggregates (FA; Cases I and II) \\
 & & \quad hierarchical aggregate (HA; Case III) \\

Porosity $\mathcal{P}$ (\texttt{ADDA} CP) & 7 &
$0.30$, $0.40$, $0.50$, $0.60$, $0.70$, $0.80$, $0.90$ \\

Porosity $\mathcal{P}$ (\texttt{MSTM} aggregates) & 3 &
$0.94$, $0.98$, $0.99$ \\

Particle size $a$ [$\mu\mathrm{m}$] & 45 &
logarithmically spaced from $0.1$ to $1000$ \\

Wavelengths with extended \texttt{MSTM} coverage [$\mathrm{mm}$] & 4 &
$\lambda=0.87$, $1.0$, $3.0$, and $7.0$ \\
\hline
\end{tabular}
\end{table}

\section{Methodology} \label{sec:methodology}

We model two particle populations: consolidated porous particles (CP), generated by carving voids in a spherical seed while preserving a spherical envelope (Figure~\ref{fig:consolidated_porous}), and highly porous \texttt{MSTM} aggregate models, including fractal aggregates (FA) and one hierarchical aggregate (HA), with $\mathcal{P}\gtrsim 0.9$ (Figure~\ref{fig:fractal_aggregates}). The explored parameter space is listed in Table~\ref{tab:param_space}. We evaluate ten wavelengths from $0.87$ to $10\,\mathrm{mm}$, spanning the ALMA/VLA bands in Table~\ref{tab:opt_const}. This interval samples the transition from scattering-affected, often optically thick emission at shorter wavelengths to optically thin emission at $\lambda\gtrsim 3\,\mathrm{mm}$ \citep{Carrasco2019}. At $3$--$4\,\mathrm{mm}$, emission is also increasingly midplane-dominated beyond $\sim 20$--$30$\,AU \citep[e.g.,][]{Macias2021_AandA_648_A33,Garufi2025TaurusCm}.

\subsection{Optical Constants} \label{subsec:optical_constants}

We adopt the DSHARP dust composition \citep{Birnstiel2018}, consistent with the multi-component mixtures commonly used in radiative transfer models of disks \citep[e.g.,][]{Woitke2016}. By volume, the mixture includes water ice ($\sim 36\%$; \citealt{WarrenBrandt2008}), astronomical silicates ($\sim 16\%$; \citealt{Draine2003}), troilite ($\sim 4\%$), and refractory organics ($\sim 44\%$; \citealt{HenningStognienko1996}; see also \citealt{Henning2010}). We compute the complex refractive index $m(\lambda)=n(\lambda)+ik(\lambda)$ with \texttt{dsharp\_opac} \citep{Birnstiel2018}, which evaluates the effective dielectric function $\epsilon_{\mathrm{eff}}(\lambda)$ using the symmetric Bruggeman rule:
\begin{equation}
\sum_{i} f_{i} \frac{\epsilon_{i} - \epsilon_{\mathrm{eff}}}{\epsilon_{i} + 2\epsilon_{\mathrm{eff}}} = 0,
\label{eq:bru_rule}
\end{equation}
where $f_i$ and $\epsilon_i$ are the volume fraction and dielectric function of component $i$. Table~\ref{tab:opt_const} lists the resulting optical constants. The corresponding bulk density of the solid mixture is $\rho_{\mathrm{bulk}} = 1.67\,\mathrm{g\,cm^{-3}}$.

\begin{table}[ht]
\centering
\caption{DSHARP optical constants and corresponding ALMA/VLA bands.}
\label{tab:opt_const}
\begin{tabular}{ccccc}
\hline
$\lambda [\mathrm{mm}]$ & $n(\lambda)$ & $k(\lambda)$ & ALMA band & VLA band\\
\hline
0.87 & 2.299 & 0.0224 & 7 & - \\
1.0 & 2.299 & 0.0203 & 7 & - \\
1.25 & 2.299 & 0.0177 & 6 & - \\
1.6 & 2.300 & 0.0152 & 5 & - \\
2.0 & 2.300 & 0.0120 & 4 & - \\
3.0 & 2.302 & 0.0083 & 3 & - \\
4.0 & 2.302 & 0.0064 & 2 & - \\
7.0 & 2.302 & 0.0042 & 1 & Q \\
9.0 & 2.302 & 0.0036 & - & Ka \\
10.0 & 2.302 & 0.0033 & - & Ka \\
\hline
\end{tabular}
\end{table}

\subsection{Consolidated Porous Particles} \label{subsec:consolidated_porous_particles}

We compute the CP models with \texttt{ADDA} (v1.4.0; \citealt{Yurkin2011}) using the discrete dipole approximation (DDA), where each target model is discretized on a cubic lattice of polarizable point dipoles. We track discretization with $|m|kd$, where $k=2\pi/\lambda$ is the wavenumber, $d$ is the lattice spacing, and $|m|$ is the modulus of the complex refractive index. We adopt $|m|kd<0.3$, which is equivalent to at least 21 dipoles per internal wavelength - much finer than typically recommended for compact particles \citep[see][]{Yurkin2011}. Orientation averaging follows the internal \texttt{ADDA} scheme: for each run, we use 32 uniformly spaced values of the Euler angle $\alpha$ and an adaptively refined $(\beta,\gamma)$ grid with 14--242 angle pairs. This setup keeps the relative error in orientation-averaged $C_{\mathrm{ext}}$ below $0.1\%$. We model CP particles as a randomly oriented ensemble \citep{Mishchenko2017}. All remaining DDA parameters follow the standard \texttt{ADDA} configuration. Convergence and realization-to-realization variability for the CP calculations are quantified in Appendix~\ref{app:convergence}.

We generate the CP models starting from a solid spherical seed on a cubic lattice, using a $128^3$ grid for $a=0.1$--$90\,\mu\mathrm{m}$ and a $256^3$ grid for $a=100$--$1000\,\mu\mathrm{m}$. For a target porosity $\mathcal{P}$, we iteratively carve non-overlapping spherical cavities until the target porosity is reached:
\begin{equation}
\mathcal{P}
= \frac{V_{\mathrm{seed}} - V_{\mathrm{final}}}{V_{\mathrm{seed}}}
= \frac{N_{\mathrm{seed}} - N_{\mathrm{final}}}{N_{\mathrm{seed}}},
\label{eq:CP_porosity}
\end{equation}

where $V_{\mathrm{final}}$ and $N_{\mathrm{final}}$ are the volume and number of occupied lattice sites in the final model. We define the seed volume as $V_{\mathrm{seed}}=N_{\mathrm{seed}}d^3$, where $N_{\mathrm{seed}}$ is the number of occupied dipoles. Each cavity of radius $r_{\mathrm{pore}}$ is centered on a randomly selected solid dipole and accepted only if it does not overlap a previously carved pore. The procedure stops when the number of removed dipoles reaches $N_{\mathrm{remove}}=\mathcal{P}N_{\mathrm{seed}}$; in practice, this criterion preserves global connectivity and avoids detached fragments in the generated models. As a size reference, we use the radius of an equivalent-volume, compact, homogeneous sphere,
\begin{equation}
a = \left(\frac{3V_{\mathrm{solid}}}{4\pi}\right)^{1/3},
\label{eq:a_V}
\end{equation}
where $V_{\mathrm{solid}}$ is the material volume. Thus, $a$ corresponds to a compact sphere with the same solid volume, and hence the same mass for fixed material density, as the porous model. The corresponding outer radius is
\begin{equation}
R_{\mathrm{out}} = a(1-\mathcal{P})^{-1/3}.
\end{equation}

By construction, all CP models with the same $a$ contain the same solid mass, while $R_{\mathrm{out}}$ increases with porosity. The CP grid therefore contains $7$ porosities $\times$ $45$ sizes $\times$ $10$ wavelengths, or $3150$ \texttt{ADDA} simulations. Table~\ref{tab:consolidated_porous_parameters} lists $|m|kd$ for the largest models. We set the pore scale by generating a reference geometry at $a=100\,\mu\mathrm{m}$ with $r_{\mathrm{pore}}=5\,\mu\mathrm{m}$ and then scaling the same dimensionless pore pattern to other sizes so that $r_{\mathrm{pore}}\propto a$. The Python script used to generate these models is available in the Zenodo repository \citep{vargas2026zenodo}.

\begin{table*}[ht]
\centering
\caption{Outer radius $R_{\mathrm{out}}$, dipole count, and $|m|kd$ for the largest consolidated porous models ($a=1\,\mathrm{mm}$).}
\label{tab:consolidated_porous_parameters}
\begin{tabular}{ccccccc}
\hline
Porosity & Outer Radius $R_\mathrm{out}$ [mm] & Dipoles & \multicolumn{4}{c}{$|m|kd$} \\
 & & & 0.87 mm & 1 mm & 3 mm & 7 mm \\
\hline
0.3 & 1.12 & 6,211,171 & 0.147 & 0.128 & 0.043 & 0.018 \\
0.4 & 1.18 & 5,257,720 & 0.154 & 0.134 & 0.045 & 0.019 \\
0.5 & 1.25 & 4,848,926 & 0.164 & 0.143 & 0.048 & 0.020 \\
0.6 & 1.35 & 3,624,480 & 0.177 & 0.154 & 0.051 & 0.022 \\
0.7 & 1.49 & 2,804,781 & 0.195 & 0.169 & 0.056 & 0.024 \\
0.8 & 1.71 & 1,883,827 & 0.223 & 0.194 & 0.065 & 0.028 \\
0.9 & 2.15 & 1,159,148 & 0.280 & 0.244 & 0.081 & 0.035 \\
\hline
\end{tabular}
\end{table*}

\subsection{\texttt{MSTM} Aggregate Models} \label{subsec:fractal_aggregates}

For the aggregate models we use \texttt{MSTM} \citep[v4.0;][]{Mackowski2011}, a frequency-domain multiple-sphere superposition code based on the $T$-matrix method for clusters of spherical monomers. \texttt{MSTM} expands the fields around each monomer in vector spherical wave functions and, for finite clusters in free space, provides cross sections and scattering-matrix elements for fixed orientations and, via analytical orientation averaging, for randomly oriented clusters. The computational cost increases rapidly with monomer number and multipole truncation order, so, for computational tractability in our production runs, we restrict the baseline calculations to $N_{\mathrm{mon}}\leq 10{,}000$ and choose $a_{\mathrm{mon}}$ so that the baseline \texttt{MSTM} aggregate grid spans equivalent-volume radii from $a=0.1\,\mu\mathrm{m}$ to $a=100\,\mu\mathrm{m}$ at all wavelengths. We then add an extended subset up to $a=1\,\mathrm{mm}$ at $\lambda=0.87$, $1.0$, $3.0$, and $7.0\,\mathrm{mm}$.

We generate the aggregates with the \texttt{aggregate\_gen} C++ code \citep{Moteki_aggregate_gen_2019}, which implements the tunable cluster--cluster aggregation (CCA) scheme of \citet{Filippov2000} for monodisperse spherical monomers. For an aggregate composed of $N_{\mathrm{mon}}$ monomers, the radius of gyration is
\begin{equation}
a_\mathrm{g} = \sqrt{\sum_{i=1}^{N_{\mathrm{mon}}} \frac{\left| \mathbf{r}_i - \mathbf{r}_{\mathrm{CM}} \right|^2}{N_{\mathrm{mon}}}},
\label{eq:a_g}
\end{equation}
where $\mathbf{r}_i$ denotes the position of monomer $i$ and $\mathbf{r}_{\mathrm{CM}}$ the center-of-mass position. Each aggregate is characterized by the fractal dimension $D_f$ and the fractal prefactor $k_f$, related through
\begin{equation}
N_{\mathrm{mon}} = k_f \left( \frac{a_\mathrm{g}}{a_{\mathrm{mon}}}\right)^{D_f}.
\label{eq:Nmon}
\end{equation}
The corresponding radius of an equivalent-volume sphere is
\begin{equation}
a = N_{\mathrm{mon}}^{1/3} a_{\mathrm{mon}},
\label{eq:a_V_FA}
\end{equation}
and the porosity is
\begin{equation}
\mathcal{P} = 1 - N_{\mathrm{mon}} \left( \frac{a_{\mathrm{mon}}}{a_\mathrm{c}} \right)^3 = 1 - \left( \frac{a}{a_\mathrm{c}} \right)^3.
\label{eq:porosity_FA}
\end{equation}

Here, $a_{\rm c}=\sqrt{5/3}\,a_{\rm g}$ is the characteristic radius. We consider three representative aggregate morphologies, each with 8192 monomers: (i) a fractal aggregate (FA) with $D_{\rm f}=2.7$ and $k_{\rm f}=0.3$ ($\mathcal{P}=0.94$), (ii) an FA with $D_{\rm f}=2.1$ and $k_{\rm f}=0.7$ ($\mathcal{P}=0.98$), and (iii) a hierarchical aggregate (HA) with $\mathcal{P}=0.99$ \citep[e.g.,][]{Kolokolova2018}. The HA model was constructed by first generating $64$ independent sub-aggregates, each containing $128$ monomers with $D_{\rm f}=1.2$ and $k_{\rm f}=1.0$. These sub-aggregates were randomly rotated, placed at random positions in a three-dimensional Cartesian frame, and translated toward the center of mass of the full system until the first inter-aggregate contact was reached, defined by a center-to-center distance $\leq 2 a_{\mathrm{mon}}$ between monomers belonging to different sub-aggregates. The final monomer positions were then fixed, yielding a fluffy 8192-monomer hierarchical aggregate with $\mathcal{P}=0.99$. Thus, $D_{\rm f}=1.2$ characterizes the internal structure of the sub-aggregates, whereas the final Case III particle should be interpreted as a hierarchical aggregate rather than as a single-scale fractal aggregate. The resulting aggregate models are shown in Figure~\ref{fig:fractal_aggregates}.

\begin{table}[ht]
\centering
\caption{Monomer size parameter, $x_{\mathrm{mon}} = 2\pi a_{\mathrm{mon}}/\lambda$, for representative \texttt{MSTM} aggregate models.}
\label{tab:monomer_size_FA}
\begin{tabular}{cccccc}
\hline
$a$ [$\mu\mathrm{m}$] & $a_{\mathrm{mon}}$ [$\mu\mathrm{m}$] & \multicolumn{4}{c}{$x_{\mathrm{mon}}$} \\
 & & 0.87 mm & 1 mm & 3 mm & 7 mm \\
\hline
1     & 0.05  & 0.0004 & 0.0003 & 0.0001 & 0.0001 \\
10    & 0.5  & 0.0036 & 0.0031 & 0.0010 & 0.0004 \\
100   & 5.0  & 0.0361 & 0.0314 & 0.0105 & 0.0045 \\
1000  & 50   & 0.3611 & 0.3142 & 0.1047 & 0.0449 \\
\hline
\end{tabular}
\end{table}

\begin{figure*}[ht!]
\centering
\includegraphics[width=17cm]{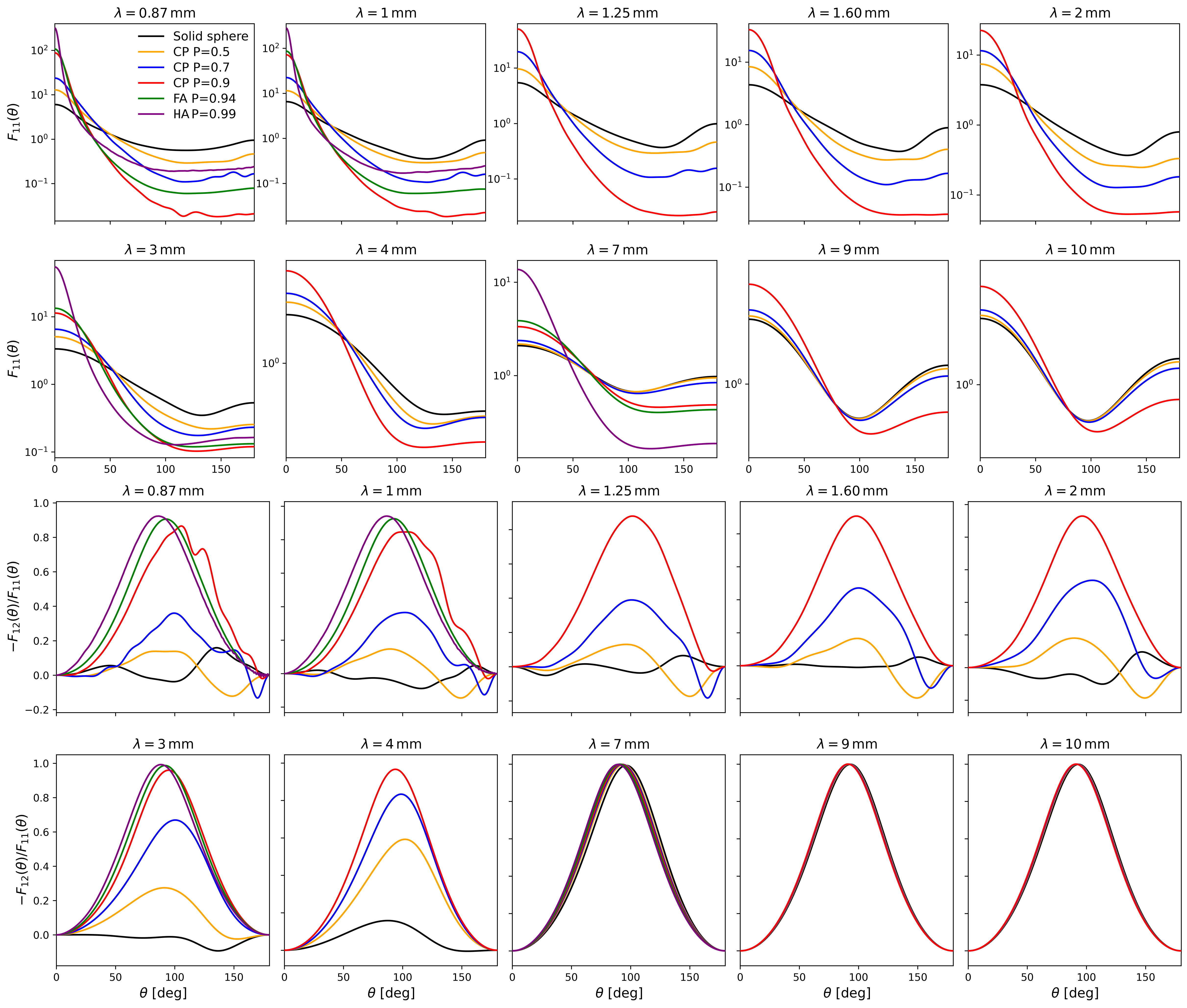}
\caption{Size-averaged scattering properties for the adopted size distribution. The upper rows show the phase function $F_{11}(\theta)$ and the lower rows show the linear polarization fraction $-F_{12}(\theta)/F_{11}(\theta)$ from $\lambda=0.87$ to $10\,\mathrm{mm}$ for compact spheres (CS), consolidated porous particles (CP; $\mathcal{P}=0.50$, $0.70$, and $0.90$), and \texttt{MSTM} aggregate models (FA with $\mathcal{P}=0.94$ and HA with $\mathcal{P}=0.99$). For the \texttt{MSTM} aggregate models, results up to $a_{\max}=1\,\mathrm{mm}$ are available only at $\lambda=0.87$, $1.0$, $3.0$, and $7.0\,\mathrm{mm}$.}
\label{fig:plot3}
\end{figure*}

For each morphology, we generate one aggregate with fixed $N_{\mathrm{mon}}$ and construction parameters, and obtain 45 sizes by scaling $a_{\mathrm{mon}}$ (and thus $a$). The baseline \texttt{MSTM} aggregate grid therefore includes $3$ morphologies $\times$ $45$ sizes $\times$ $10$ wavelengths, i.e., 1350 \texttt{MSTM} simulations. The additional calculations up to $a=1\,\mathrm{mm}$ at $\lambda=0.87$, $1.0$, $3.0$, and $7.0\,\mathrm{mm}$ form the extended subset beyond this baseline grid. To assess the impact of using a single aggregate realization, realization-to-realization scatter for a representative $D_f=2.1$ FA model is quantified in Appendix~\ref{app:mstm_realizations}. To compare directly with CP models, we select monomer radii comparable to characteristic CP pore sizes. Table~\ref{tab:monomer_size_FA} lists the adopted monomer radii and monomer size parameters for representative models with $a=1\,\mu\mathrm{m}$, $10\,\mu\mathrm{m}$, $100\,\mu\mathrm{m}$, and $1000\,\mu\mathrm{m}$ at four wavelengths; the $a=1000\,\mu\mathrm{m}$ row corresponds to the extended \texttt{MSTM} aggregate subset. In the main figures, we show only the $\mathcal{P}=0.94$ FA and $\mathcal{P}=0.99$ HA cases for readability.

\section{Results and Discussion} \label{sec:Results}

\subsection{Phase Function and Linear Polarization Fraction}\label{subsec:results_phase}

The upper rows of Figure~\ref{fig:plot3} show the size-averaged phase function $F_{11}(\theta)$. Across the studied spectral range, increasing porosity systematically sharpens the forward cusp, as reflected in larger $F_{11}$ values at $\theta\lesssim 10^\circ$, while generally weakening the backscattering lobe. This ordering is clearest at intermediate wavelengths ($\lambda=2$--$4\,\mathrm{mm}$). At the longest wavelengths ($\lambda\ge 7\,\mathrm{mm}$), the phase functions become smoother and exhibit a broad minimum near $\theta\simeq 90^\circ$.

The lower rows of Figure~\ref{fig:plot3} show the linear polarization fraction $P(\theta)$. At short wavelengths, compact spheres and low-porosity particles show a limited positive peak and a pronounced negative branch at large scattering angles ($\theta\gtrsim 120^\circ$), most clearly for $\mathcal{P}\lesssim 0.7$. Increasing porosity strengthens the peak, shifts it toward $\theta\simeq 90^\circ$, and progressively suppresses the negative branch; the highest-porosity case remains positive at nearly all scattering angles. At intermediate wavelengths ($\lambda=2$--$4\,\mathrm{mm}$), the polarization curves are smoother and largely positive, with a broad maximum near $\theta\sim 90^\circ$. By $\lambda\ge 7\,\mathrm{mm}$, the curves approach the Rayleigh regime.

\begin{figure*}[ht!]
\centering
\includegraphics[width=17cm]{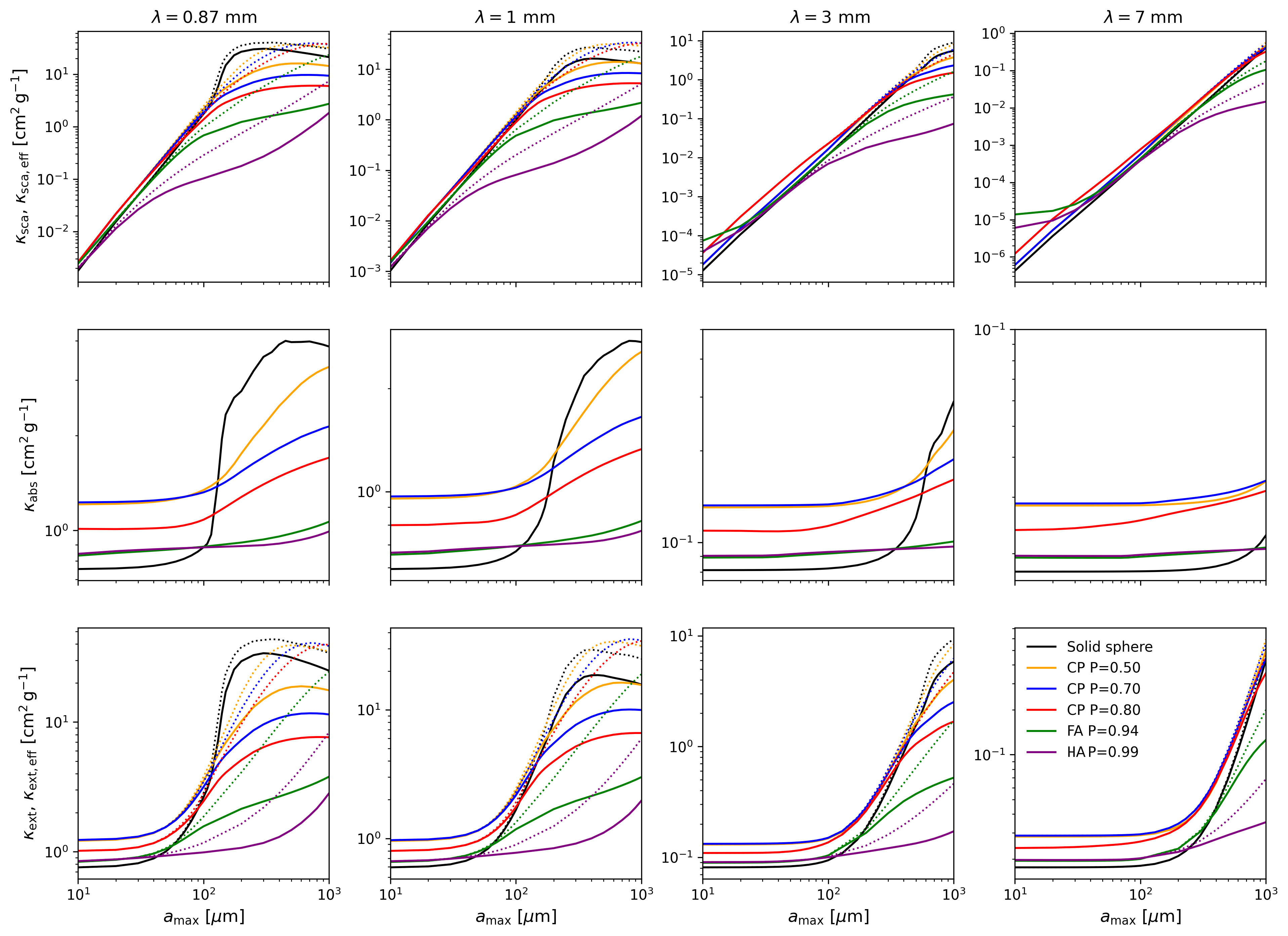}
\caption{Size-averaged mass opacities for a size distribution with $a_{\min}=0.1\,\mu\mathrm{m}$, shown as functions of the maximum particle size $a_{\max}$ at four representative wavelengths, $\lambda=0.87$, $1.0$, $3.0$, and $7.0\,\mathrm{mm}$. Top panels: scattering opacity $\kappa_{\mathrm{sca}}$ (dotted) together with the effective scattering opacity $\kappa_{\mathrm{sca,eff}}$ (solid). Middle panels: absorption opacity $\kappa_{\mathrm{abs}}$. Bottom panels: extinction opacity $\kappa_{\mathrm{ext}}$ (dotted) together with $\kappa_{\mathrm{ext,eff}}$ (solid). Curves compare compact spheres (CS), consolidated porous particles (CP) with $\mathcal{P}=0.50$, $0.70$, and $0.90$, and \texttt{MSTM} aggregate models, including FA with $\mathcal{P}=0.94$ and HA with $\mathcal{P}=0.99$.}
\label{fig:plot4}
\end{figure*}

At all wavelengths, the \texttt{MSTM} aggregate models retain a Rayleigh-like linear-polarization profile. This behavior can be understood from the monomer size parameters listed in Table~\ref{tab:monomer_size_FA}. Because these aggregates are highly porous and their monomers remain relatively small compared with the wavelength, the aggregate response is expected to be closer to the Rayleigh--Debye--Gans (RDG) regime, in which scattering is described by weakly coupled monomer contributions, while the large-scale aggregate structure mainly modulates the angular intensity pattern \citep{Farias1996}.

\subsection{Mass Opacities}\label{subsec:results_opacity}

\begin{figure*}[ht!]
\centering
\includegraphics[width=15cm]{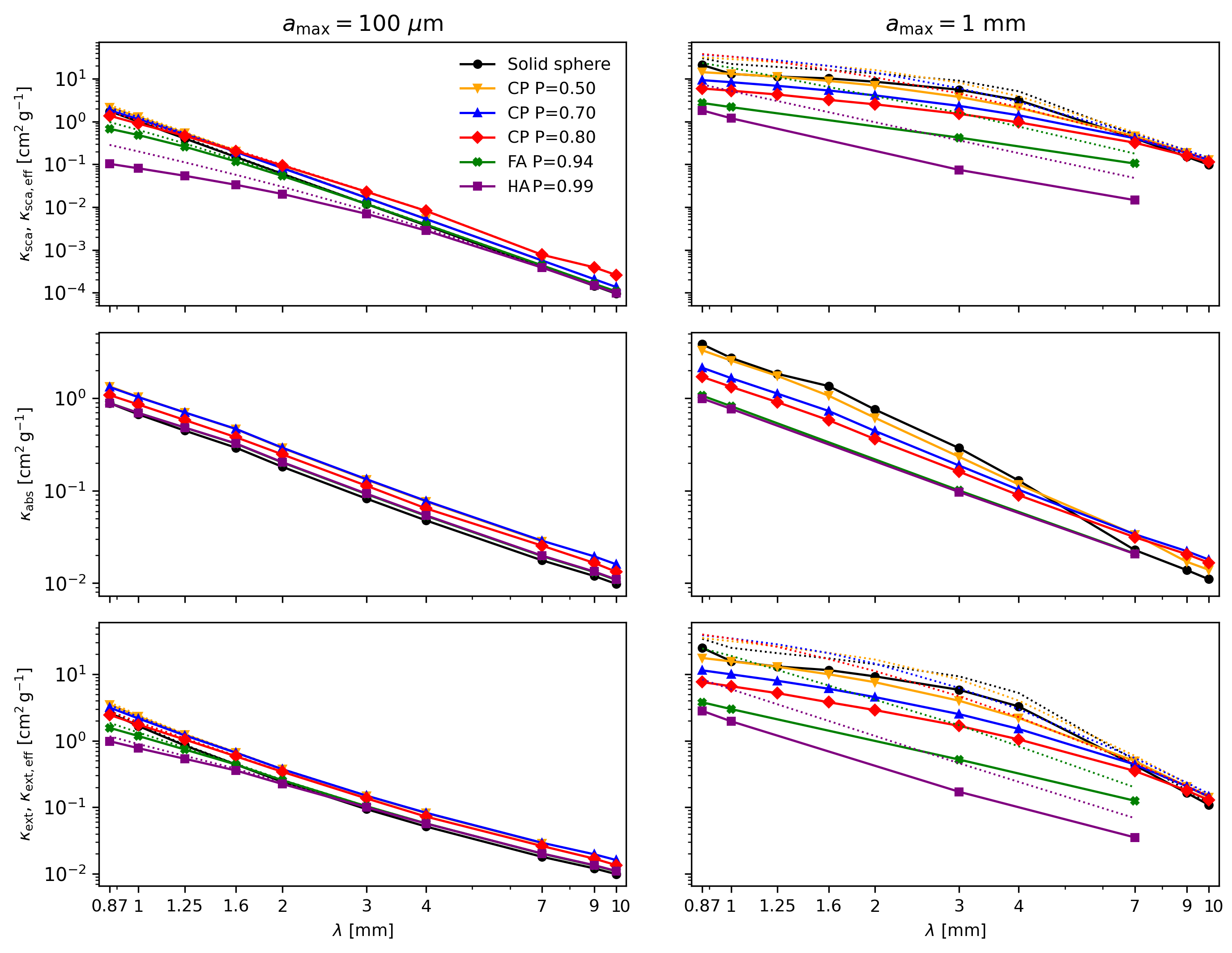}
\caption{Spectral behavior of the size-averaged mass opacities for a size distribution with $a_{\min}=0.1\,\mu\mathrm{m}$. Left and right columns correspond to $a_{\max}=100\,\mu\mathrm{m}$ and $a_{\max}=1\,\mathrm{mm}$, respectively. Top panels: scattering opacity $\kappa_{\mathrm{sca}}$ (dotted) together with the effective scattering opacity $\kappa_{\mathrm{sca,eff}}$ (solid). Middle panels: absorption opacity $\kappa_{\mathrm{abs}}$. Bottom panels: extinction opacity $\kappa_{\mathrm{ext}}$ (dotted) together with $\kappa_{\mathrm{ext,eff}}$ (solid). The same cases as in Figure~\ref{fig:plot4} are shown. \texttt{MSTM} aggregate results are shown only for the wavelength subset $\lambda=0.87$, $1.0$, $3.0$, and $7.0\,\mathrm{mm}$.}
\label{fig:plot5}
\end{figure*}

Figure~\ref{fig:plot4} summarizes size-averaged absorption, scattering, and extinction opacities as functions of $a_{\max}$ at four representative wavelengths. For small $a_{\max}$, absorption dominates extinction. Once $a_{\max}$ enters the Mie/interference regime, both $\kappa_{\mathrm{sca}}$ and $\kappa_{\mathrm{ext}}$ rise rapidly and then approach a plateau. Compact spheres and low-porosity CP models reach the highest values, whereas increasing porosity systematically reduces opacities, most strongly for \texttt{MSTM} aggregate models. The anisotropy correction also grows with size and porosity: the gap between $\kappa_{\mathrm{sca}}$ and $\kappa_{\mathrm{sca,eff}}$ increases as $g$ increases. At fixed dust mass, $\kappa_{\mathrm{abs}}$ decreases with porosity, and \texttt{MSTM} aggregate models give the lowest values once $a_{\max}\gtrsim 100\,\mu\mathrm{m}$. Figure~\ref{fig:plot5} shows the wavelength dependence of the size-averaged mass opacities at fixed $a_{\max}$. For $a_{\max}=100\,\mu\mathrm{m}$, $\kappa_{\mathrm{sca}}$, $\kappa_{\mathrm{abs}}$, and $\kappa_{\mathrm{ext}}$ decrease smoothly with $\lambda$, and $\kappa_{\mathrm{sca,eff}}\approx\kappa_{\mathrm{sca}}$, indicating weak forward peaking. For $a_{\max}=1\,\mathrm{mm}$, compact spheres and low-porosity CP models yield the largest $\kappa_{\mathrm{sca}}$ and $\kappa_{\mathrm{ext}}$. At longer wavelengths, all curves converge as the size parameter decreases.

An observational consequence of adopting porous-particle opacities is a systematic change in the disk properties inferred from continuum emission. In the optically thin limit, $F_\nu \propto \kappa_{\mathrm{abs}}(\nu)\,M_{\mathrm{dust}}\,B_\nu(T_{\mathrm{dust}})$ \citep[see Section~3 of][]{Carrasco2019}; therefore, for the opacity trends found here, compact-sphere opacities would imply smaller $M_{\mathrm{dust}}$ than porous-particle opacities. In the optically thin Rayleigh--Jeans limit, the continuum spectral index can be approximated as $\alpha \simeq 2+\beta_\kappa$, with $\kappa_{\mathrm{abs}}(\nu)\propto\nu^{\beta_\kappa}$. Thus, porosity-induced changes in the opacity slope alter the mapping between an observed spectral index and the inferred $a_{\max}$. A full comparison with spectral energy distribution (SED) constraints requires radiative-transfer modeling, but the opacity trends shown here indicate that porosity can shift the particle sizes inferred from continuum slopes. In optically thick regions, neglecting scattering and treating extinction as pure absorption can bias the inferred dust mass low for a fixed temperature and observed intensity \citep{Carrasco2019,Sierra2020ApJ}.

\begin{figure*}[ht!]
\centering
\includegraphics[width=15cm]{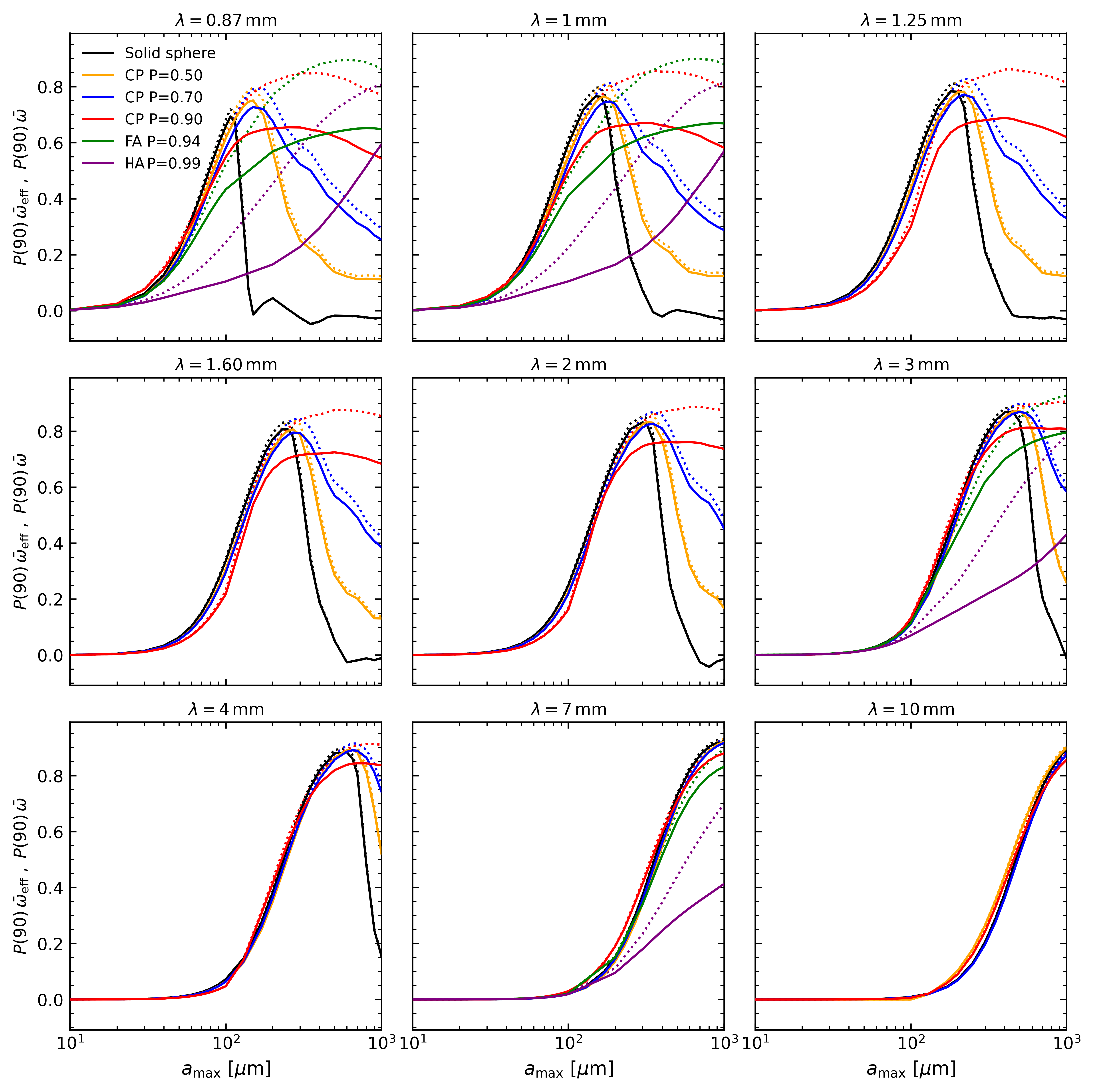}
\caption{$P(90^\circ)\,\bar{\omega}$ (dotted) and $P(90^\circ)\,\bar{\omega}_{\mathrm{eff}}$ (solid) as functions of the maximum particle size $a_{\max}$ for a power-law size distribution, $n(a)\propto a^{-3.5}$, with $a_{\min}=0.1\,\mu\mathrm{m}$. The panels span $\lambda=0.87$--$10\,\mathrm{mm}$. For the \texttt{MSTM} aggregate models, calculations up to $a_{\max}=1\,\mathrm{mm}$ are available only at $\lambda=0.87$, $1.0$, $3.0$, and $7.0\,\mathrm{mm}$.}
\label{fig:plot6}
\end{figure*}

\subsection{Polarized Scattered Emission}
\label{subsec:resultsP_albedo}

Figure~\ref{fig:plot6} shows $P(90^\circ)\,\bar{\omega}_{\mathrm{eff}}$ as a function of $a_{\max}$. For compact spheres, this quantity reaches a narrow maximum when the largest grains approach the Mie/interference scale, $a_{\max}\sim\lambda/2\pi$. At shorter wavelengths, this maximum is followed by a sign reversal in $P(90^\circ)$, which becomes negative beyond $\sim200\,\mu\mathrm{m}$ at $\lambda=0.87\,\mathrm{mm}$, $\sim400\,\mu\mathrm{m}$ at $1.0\,\mathrm{mm}$, and $\sim1\,\mathrm{mm}$ at $3.0\,\mathrm{mm}$. For $\lambda\ge 7\,\mathrm{mm}$, the corresponding Mie/interference scale lies outside the explored range, beyond $a_{\max}=1\,\mathrm{mm}$. For CP particles with $\mathcal{P}\lesssim 0.70$, the maximum shifts to larger $a_{\max}$, and highly porous particles show a broader maximum and maintain high positive values up to $a_{\max}=1\,\mathrm{mm}$. At short wavelengths and $a_{\max}=1\,\mathrm{mm}$, $P(90^\circ)\,\bar{\omega}_{\mathrm{eff}}$ remains low for compact spheres and low-porosity CP models, whereas it increases strongly for high-porosity models. At intermediate wavelengths, the asymmetry correction is modest at low porosity but substantial at high porosity, where $P(90^\circ)\,\bar{\omega}_{\mathrm{eff}}$ remains well below $P(90^\circ)\,\bar{\omega}$. At $\lambda=7\,\mathrm{mm}$, all models increase as $a_{\max}$ approaches $\lambda/2\pi \simeq 1\,\mathrm{mm}$.

\begin{figure*}[ht!]
\centering
\includegraphics[width=15cm]{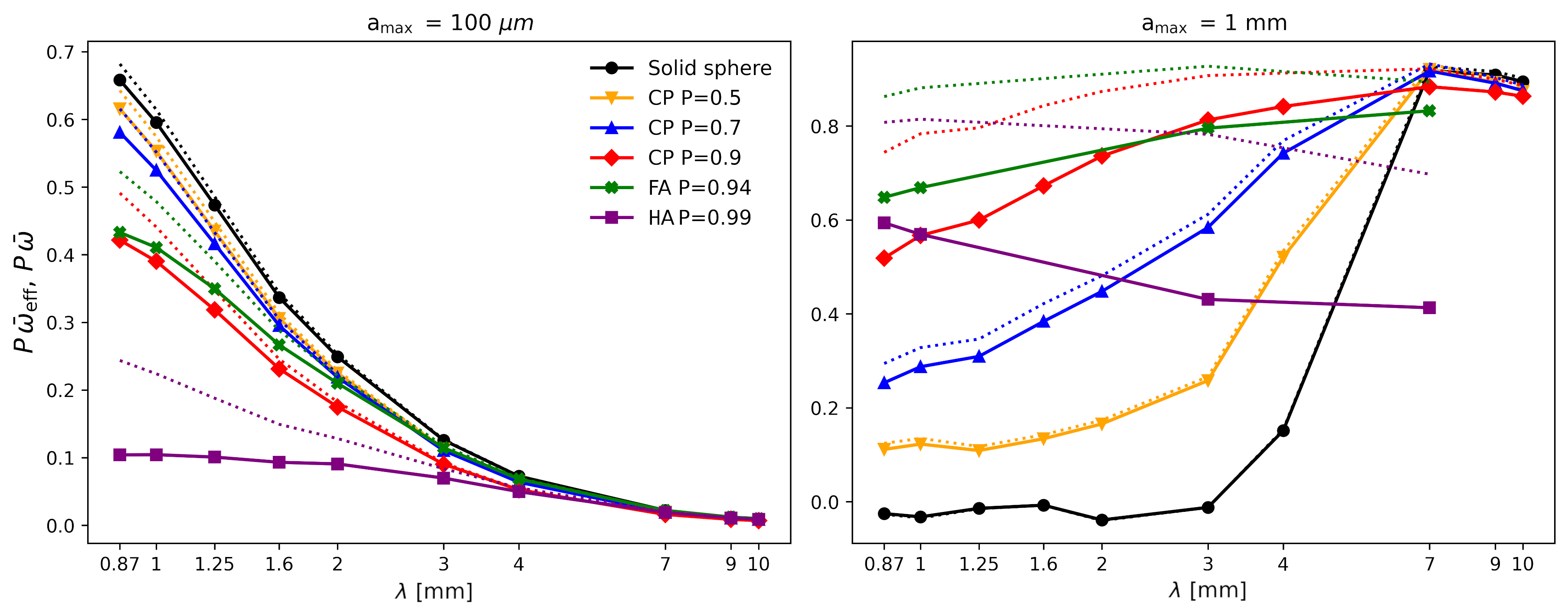}
\caption{Spectral behavior of $P(90^\circ)\,\bar{\omega}$ (dotted) and $P(90^\circ)\,\bar{\omega}_{\mathrm{eff}}$ (solid) for a size distribution with $a_{\min}=0.1\,\mu\mathrm{m}$ and $a_{\max}=100\,\mu\mathrm{m}$ (left) or $1\,\mathrm{mm}$ (right). The same particle morphologies as in Figure~\ref{fig:plot6} are shown.}
\label{fig:plot7}
\end{figure*}

\begin{figure}[ht!]
\includegraphics[width=8cm]{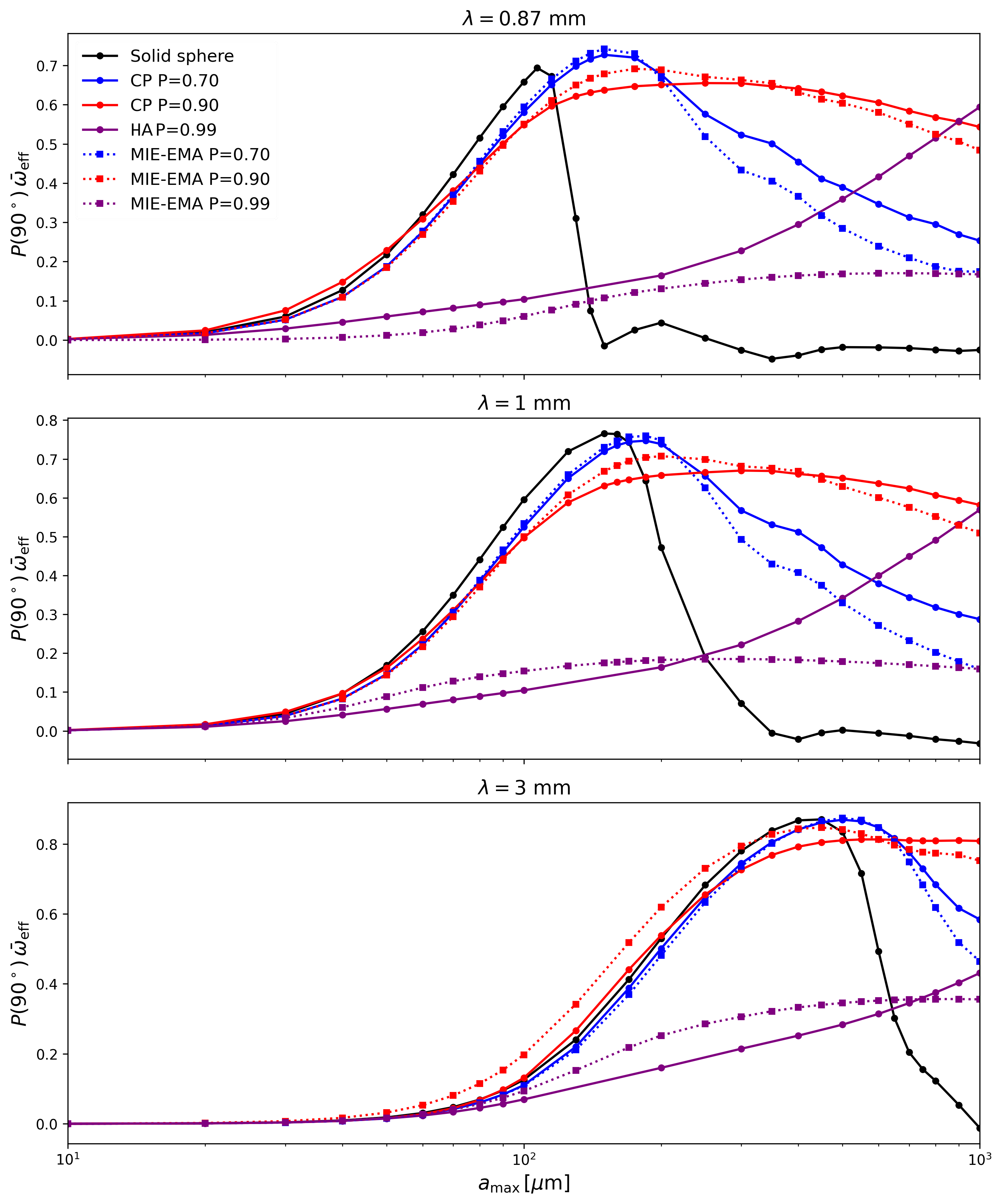}
\caption{$P(90^\circ)\,\bar{\omega}_{\mathrm{eff}}$ for $\lambda=0.87\,\mathrm{mm}$ (top), $1.0\,\mathrm{mm}$ (middle), and $3.0\,\mathrm{mm}$ (bottom). Solid lines show the results for compact spheres and porous particles: CP with $\mathcal{P}=0.70$ and $0.90$, and HA with $\mathcal{P}=0.99$. Dotted lines show the Mie--EMA models for $\mathcal{P}=0.70$, $0.90$, and $0.99$, obtained by applying the Bruggeman rule to the DSHARP composition.}
\label{fig:plot8}
\end{figure}

Figure~\ref{fig:plot7} shows the spectral dependence of $P(90^\circ)\,\bar{\omega}$ and $P(90^\circ)\,\bar{\omega}_{\mathrm{eff}}$ for $a_{\max}=100\,\mu\mathrm{m}$ and $1\,\mathrm{mm}$. For $a_{\max}=100\,\mu\mathrm{m}$, all morphologies decrease monotonically as the size parameter decreases toward the Rayleigh regime. For most models, $\bar{\omega}_{\mathrm{eff}}$ closely follows $\bar{\omega}$, with the largest reductions for $\mathcal{P}=0.94$--$0.99$. For $a_{\max}=1\,\mathrm{mm}$, compact spheres produce weak polarized scattered emission at $\lambda=0.87$--$3\,\mathrm{mm}$, whereas porous particles and aggregates maintain comparatively high values of $P(90^\circ)\,\bar{\omega}_{\mathrm{eff}}$. At $\lambda\ge 7\,\mathrm{mm}$, most models show an increase in $P(90^\circ)\,\bar{\omega}_{\mathrm{eff}}$ as $a_{\max}$ approaches the Mie/interference scale, $a_{\max}\sim\lambda/2\pi$. The main exception is the hierarchical aggregate, for which $P(90^\circ)\,\bar{\omega}_{\mathrm{eff}}$ remains significantly lower. Compact spheres produce a narrow peak in $P(90^\circ)\,\bar{\omega}_{\mathrm{eff}}$ near the Mie/interference regime, whereas porous particles maintain comparatively high values over a broader wavelength range and show a largely monotonic dependence on $\mathcal{P}$. This behavior suggests that sufficiently sensitive multi-wavelength polarization measurements may help constrain particle porosity.

Our results indicate that porosity broadens the range of particle sizes capable of producing significant polarized self-scattering, rather than confining it to the compact-sphere condition $a_{\max}\sim\lambda/2\pi$. In disks such as HL~Tau, where compact-sphere models give $a_{\max}\sim 160\,\mu\mathrm{m}$ in ALMA Bands~7 and~6 \citep{Kataoka2015}, porous particles can sustain comparable $P(90^\circ)\,\bar{\omega}_{\mathrm{eff}}$ up to $a_{\max}\sim 1\,\mathrm{mm}$. Porosity also changes $\kappa_{\mathrm{abs}}(\lambda)$ at fixed $a_{\max}$, especially near $1\,\mathrm{mm}$, so a given continuum slope can map to different particle sizes. Robust constraints therefore require multi-wavelength radiative-transfer modeling with porosity-dependent opacities.

\subsection{Comparison with Previous Studies}\label{subsec:comparison}
\citet{KirchschlagerBertrang2020} modeled self-scattering polarization by aligned spheroids for $n(a)\propto a^{-3.5}$. Their peak $P(90^\circ)\,\bar{\omega}$ values are comparable to those in Figure~\ref{fig:plot6}, but their polarization window is narrower (see their Figure 6). In our models, high porosity sustains similar amplitudes over larger sizes. \citet{Zhang2023} used Mie theory plus EMA, with the Bruggeman mixing rule, to evaluate $P(90^\circ)\,\bar{\omega}_{\mathrm{eff}}$ over a wide size range. To suppress large size-dependent oscillations, they smoothed the single-size polarization by averaging $Z_{11}(90^\circ)$ and $Z_{12}(90^\circ)$ over neighboring sizes before computing $P(90^\circ)$ (their Appendix~E and Figure~14). Because our explicit porous particle calculations use a coarser size grid, we computed additional Mie--EMA models on the same 45 sizes used in the \texttt{ADDA}/\texttt{MSTM} runs for a direct comparison. We mixed the DSHARP composition with vacuum using the Bruggeman rule for $\mathcal{P}=0.70$, $0.90$, and $0.99$ and then computed the corresponding Mie solutions. The resulting curves are shown in Figure~\ref{fig:plot8}. Within the common size range, our porous particles yield slightly higher $P(90^\circ)\,\bar{\omega}_{\mathrm{eff}}$ than the corresponding Mie--EMA models. Thus, porosity is represented structurally while retaining the DSHARP optical constants for the solid material, rather than being approximated through an effective refractive index as in Mie--EMA.

According to \citet{Wolff1994,Wolff1998}, effective-medium models become inaccurate once pores or inclusions leave the Rayleigh regime. In our baseline CP models, $r_{\mathrm{pore}}\simeq 5$--$50\,\mu\mathrm{m}$; at $\lambda_{\min}=0.87\,\mathrm{mm}$, this implies $2\pi r_{\mathrm{pore}}/\lambda\lesssim 0.4$. EMA deviations should therefore grow for larger particles and less subwavelength pores. The clearest breakdown appears at very high porosity: the Mie--EMA curve for $\mathcal{P}=0.99$ is nearly flat with $a_{\max}$, while the \texttt{MSTM} HA curve increases monotonically (Figure~\ref{fig:plot8}), consistent with the known limitations of representing very porous aggregate structures with a homogeneous effective medium \citep[e.g.,][]{Tazaki2016ApJ823_70}. For computational feasibility, our \texttt{MSTM} aggregate models adopt monomer sizes that are larger than the $\sim 400\,\mathrm{nm}$ values inferred from infrared constraints \citep[e.g.,][]{TazakiDominik2022}. In Appendix~\ref{app:pore_mon_size}, we show that the CP results are only weakly affected by the tested pore sizes, whereas the \texttt{MSTM} aggregate results are more sensitive to the adopted monomer size: decreasing $a_{\mathrm{mon}}$ lowers $\bar{\omega}_{\mathrm{eff}}$, especially at $\lambda=0.87$--$1.0$ mm. The high-$a_{\max}$ \texttt{MSTM} aggregate results should therefore be interpreted as dependent on the computationally feasible monomer sizes used here.
\section{Summary and Conclusions} \label{sec:Conclusions}
In this work, we modeled the scattering properties of two porous dust populations across ten wavelengths from $0.87$ to $10\,\mathrm{mm}$. Relative to compact spheres, increasing porosity strengthens forward scattering, increases $g$, and lowers $\bar{\omega}_{\mathrm{eff}}$. Porosity suppresses Mie/interference oscillations, reduces $\kappa_{\mathrm{abs}}$ at millimeter sizes, and increases the relative contribution of $\kappa_{\mathrm{sca,eff}}$ to effective extinction. Consequently, porous-particle opacities imply larger dust masses than compact-sphere opacities for the same continuum flux.

Compact spheres exhibit higher $P(90^\circ)\,\bar{\omega}_{\mathrm{eff}}$ values around $a_{\max}\sim \lambda/2\pi$, whereas porous particles maintain comparatively high values over a broader size interval, extending up to $a_{\max}\sim 1\,\mathrm{mm}$. Porosity therefore makes the inference of $a_{\max}$ from continuum polarization morphology-dependent and can help reconcile the $\sim 160\,\mu\mathrm{m}$ sizes inferred for HL~Tau from polarization \citep{Kataoka2015} with millimeter-sized particles implied by continuum emission constraints \citep{Carrasco2019}. 

Our results show that porosity alters the polarized continuum spectrum relative to compact spheres.
Multi-wavelength polarization observations across the $\sim1$--$10\,\mathrm{mm}$ range can constrain particle porosity if the sensitivity to low polarization fractions and the angular resolution are sufficient to separate disk regions. Such studies are currently most accessible near $1\,\mathrm{mm}$ with ALMA; observations around $3\,\mathrm{mm}$ are feasible but high-resolution mapping is more demanding. At longer wavelengths, the continuum and polarized signals become fainter, but the ALMA Wideband Sensitivity Upgrade and the ngVLA should improve porosity-sensitive diagnostics, with complementary constraints from the SKA at centimeter wavelengths.

Compared with our explicit porous-particle models, Mie--EMA captures the broader polarization window at intermediate porosity, but it deviates more at very high porosity, especially for the very high-porosity HA case. This highlights the limitations of EMA outside the Rayleigh regime. Because the monomer-size tests show that decreasing $a_{\mathrm{mon}}$ lowers $\bar{\omega}_{\mathrm{eff}}$, especially at shorter millimeter wavelengths, the high-$a_{\max}$ \texttt{MSTM} aggregate results should be interpreted as an upper-limit case for the adopted computationally feasible monomer sizes.

Future work will focus on exploring compositions in which refractory organics are replaced by more opaque amorphous-carbon components, in line with the opacity prescriptions used by \citet{Ricci2010} and in the DIANA framework of \citet{Woitke2016}. It will also extend the catalog to $a_{\max}\sim 1\,\mathrm{cm}$ and incorporate the resulting scattering matrices and opacities into full 3D radiative transfer simulations of protoplanetary disks.
\begin{acknowledgments}
The authors acknowledge the anonymous referee for valuable comments and suggestions that helped strengthen the interpretation of the results. This work was supported by grant PID2024-156713OB-I00, and by the Severo Ochoa grant CEX2021-001131-S, all funded by MCIN/AEI/10.13039/501100011033. The authors also acknowledge the use of the computational facilities of the Instituto de Astrof\'{\i}sica de Andaluc\'{\i}a (IAA-CSIC), computing time on the MIZTLI supercomputer of the Universidad Nacional Aut\'onoma de M\'exico (UNAM) through DGTIC project LANCAD-UNAM-DGTIC-446, and the computing resources of CRIANN (Normandy, France). G.V. acknowledges support from grant PRE2022-103488, funded by MCIN/AEI/10.13039/501100011033. C.C.-G. and J.M.J.-D. acknowledge support from UNAM DGAPA-PAPIIT grant IG101224. M.Y. acknowledges support from the Normandy Region through project RADDAERO. J.M.J.-D. gratefully acknowledges financial support from SECIHTI through the Estancias Posdoctorales por M\'exico program, CVU 864232.
\end{acknowledgments}
\begin{contribution}
G.~Vargas: writing---original draft, writing---review and editing, computational implementation, methodology, formal analysis, validation. D.~Guirado: writing---review and editing, methodology, model development, computational implementation, formal analysis, results interpretation. C.~Carrasco-Gonz\'alez: writing---review and editing, methodology, model development, computational implementation, formal analysis, results interpretation. O.~Mu\~noz: writing---review and editing, methodology, model development, computational implementation, formal analysis, results interpretation. M.~A.~Yurkin: writing---review and editing, software (ADDA), numerical accuracy, model development, results interpretation. E.~Mac\'{\i}as: writing---review and editing, conceptualization, validation, results interpretation. J.~M.~J\'aquez-Dom\'{\i}nguez: writing---review and editing, results interpretation. F.~J.~Garc\'{\i}a-Izquierdo: writing---review and editing, interpretation of laboratory measurements, results interpretation.
\end{contribution}

\section*{Data Availability}

The particle catalog and supporting scripts used in this work are archived in the Zenodo repository as \textit{PSCat: Porous Particles Scattering Catalog}, version 1.0 \citep{vargas2026zenodo}, DOI: 10.5281/zenodo.19485482.
\software{ADDA \citep{Yurkin2011},
MSTM \citep{Mackowski2011},
dsharp\_opac \citep{Birnstiel2018},
aggregate\_gen \citep{Moteki_aggregate_gen_2019},
NumPy \citep{Harris2020NumPy},
Matplotlib \citep{Hunter2007Matplotlib},
pathlib \citep{PEP428}}

\appendix

\restartappendixnumbering
\section{Accuracy and Realization Test of the ADDA Models}
\label{app:convergence}

We tested the accuracy of the adopted \texttt{ADDA} numerical setup at $\lambda=1\,\mathrm{mm}$ for CP particles with $\mathcal{P}=0.80$ and three representative equivalent-volume radii, $a=60$, $160$, and $1000\,\mu\mathrm{m}$. For each size, the adopted single-realization calculation was compared with a reference solution obtained by averaging five independent porous-particle realizations computed with finer numerical settings, doubling $n_\alpha$, $n_\beta$, $n_\gamma$, and $N_{\mathrm{grid}}$. Table~\ref{tab:appendixA_accuracy} summarizes the differences for the phase function, the linear polarization fraction, the extinction, absorption, and scattering cross sections, and the effective albedo. These differences therefore include both the change in numerical settings and realization-to-realization scatter. For angular quantities, $\Delta F_{11}$ denotes the RMS relative difference over scattering angle, while $\Delta P$ denotes the RMS absolute difference in percentage points. For integrated quantities, $\Delta C_{\mathrm{ext}}$, $\Delta C_{\mathrm{abs}}$, $\Delta C_{\mathrm{sca}}$, and $\Delta\bar{\omega}_{\mathrm{eff}}$ are relative differences with respect to the reference mean.

\begin{table}[ht]
\centering
\caption{Differences between the adopted single-realization \texttt{ADDA} setup and the reference mean at $\lambda=1\,\mathrm{mm}$.}
\label{tab:appendixA_accuracy}
\setlength{\tabcolsep}{3.5pt}
\begin{tabular}{ccccccc}
\hline
$a$ 
& $\Delta F_{11}$ 
& $\Delta P$ 
& $\Delta C_{\mathrm{ext}}$ 
& $\Delta C_{\mathrm{abs}}$ 
& $\Delta C_{\mathrm{sca}}$ 
& $\Delta \bar{\omega}_{\mathrm{eff}}$ \\
$[\mu\mathrm{m}]$ 
& $[\%]$ 
& [pp] 
& $[\%]$ 
& $[\%]$ 
& $[\%]$ 
& $[\%]$ \\
\hline
60   & 2.3  & 0.49 & 3.3 & 1.2 & 6.0 & 3.4 \\
160  & 2.1  & 0.87 & 8.6 & 2.1 & 9.5 & 1.8 \\
1000 & 13.8 & 13.5 & 3.3 & 5.6 & 3.2 & 0.5 \\
\hline
\end{tabular}
\end{table}

The adopted setup reproduces the reference solution at the few-percent level for $a=60$ and $160\,\mu\mathrm{m}$ in the angular scattering curves, and within $\sim 10\%$ in the integrated cross sections. For the largest particle, $a=1000\,\mu\mathrm{m}$, the angular differences are larger because the scattering matrix develops a more oscillatory structure, making the result more sensitive to numerical resolution and orientation averaging. Nevertheless, the corresponding differences in $C_{\mathrm{ext}}$, $C_{\mathrm{abs}}$, and $C_{\mathrm{sca}}$ remain below $10\%$, while the effective albedo differs by only $0.5\%$. These tests therefore indicate that the adopted numerical setup is adequate for the trends discussed in the main text, with the largest angular uncertainty occurring for the largest particle.
\restartappendixnumbering
\section{Realization Test of the MSTM Models}
\label{app:mstm_realizations}

We quantified realization-to-realization scatter in the \texttt{MSTM} calculations for the $D_f=2.1$ fractal aggregates. We generated five independent aggregates with the same construction scheme, fixed $a=1000\,\mu\mathrm{m}$ and $N_{\mathrm{mon}}=8192$, and repeated the calculations at $\lambda=0.87$, $1.0$, $3.0$, and $7.0\,\mathrm{mm}$. All five realizations have the same global compactness and belong to the same nominal $D_f\sim2$ family, so this test probes structural realization scatter rather than numerical convergence.

Table~\ref{tab:appendixB_mstm_scatter} summarizes the scatter across the five realizations. The integrated quantities show weak realization scatter: below $1.2\%$ for $C_{\mathrm{ext}}$ and $C_{\mathrm{sca}}$, below $0.7\%$ for $C_{\mathrm{abs}}$, and below $0.14\%$ for $\bar{\omega}_{\mathrm{eff}}$. The angular curves show larger realization-to-realization variations in the fine structure of the normalized phase function, with mean RMS scatter of $5.1$--$5.5\%$ and mean maximum deviations of $24$--$26\%$. In contrast, the polarization curve remains more stable, with mean RMS absolute scatter of $(4.6$--$5.3)\times10^{-3}$. For this representative $D_f=2.1$ FA case, the main trends discussed in the paper are therefore controlled primarily by wavelength rather than by the specific aggregate realization. For the adopted monomer size and construction scheme, the realization-to-realization scatter remains small compared with the wavelength-dependent changes. This supports the use of one representative realization in the corresponding FA grid.

\begin{table}[ht]
\centering
\caption{Realization scatter for the $D_f=2.1$ \texttt{MSTM} fractal aggregates.}
\label{tab:appendixB_mstm_scatter}
\setlength{\tabcolsep}{4.0pt}
\begin{tabular}{lc}
\hline
Quantity & Scatter across realizations \\
\hline
$C_{\mathrm{ext}}$ & $<1.2\%$ \\
$C_{\mathrm{abs}}$ & $<0.7\%$ \\
$C_{\mathrm{sca}}$ & $<1.2\%$ \\
$\bar{\omega}_{\mathrm{eff}}$ & $<0.14\%$ \\
$P(90^\circ)$ & $<1.0\%$ \\
$P(90^\circ)\,\bar{\omega}_{\mathrm{eff}}$ & $<1.0\%$ \\
$F_{11}$ RMS & $5.1$--$5.5\%$ \\
$F_{11}$ max. & $24$--$26\%$ \\
$P(\theta)$ RMS & $(4.6$--$5.3)\times10^{-3}$ \\
$P(\theta)$ max. & $(1.2$--$1.7)\times10^{-2}$ \\
\hline
\end{tabular}
\end{table}
\restartappendixnumbering
\section{Effect of Pore and Monomer Size}
\label{app:pore_mon_size}

To isolate the role of the internal pore scale in the consolidated porous (CP) models, we repeated the \texttt{ADDA} calculations for particles with fixed porosity $\mathcal{P}=0.70$ using two representative equivalent-volume radii, $a=100$ and $1000\,\mu\mathrm{m}$, at $\lambda=0.87$, $1.0$, $3.0$, and $7.0\,\mathrm{mm}$. For each $(a,\lambda)$ pair, we generated seven models with different characteristic pore radii, spanning $r_{\mathrm{pore}}=5$--$18\,\mu\mathrm{m}$ for $a=100\,\mu\mathrm{m}$ and $r_{\mathrm{pore}}=20$--$90\,\mu\mathrm{m}$ for $a=1000\,\mu\mathrm{m}$. The corresponding baseline pore sizes in the main CP grid, $r_{\mathrm{pore}}=5$ and $50\,\mu\mathrm{m}$ for $a=100$ and $1000\,\mu\mathrm{m}$, respectively, are included within these intervals. Table~\ref{tab:appendixC_pore_size} shows that the pore-size dependence is weak for $a=100\,\mu\mathrm{m}$ at all four wavelengths. For $a=1000\,\mu\mathrm{m}$, the strongest sensitivity appears at $0.87$--$1.0\,\mathrm{mm}$, mainly through changes in $P(90^\circ)\,\bar{\omega}_{\mathrm{eff}}$. At $3.0$ and $7.0\,\mathrm{mm}$, however, $\bar{\omega}_{\mathrm{eff}}$ remains nearly unchanged and the variation in $P(90^\circ)\,\bar{\omega}_{\mathrm{eff}}$ is much smaller.

\begin{table}[ht]
\centering
\caption{Pore-size test for CP models with fixed porosity $\mathcal{P}=0.70$.}
\label{tab:appendixC_pore_size}
\setlength{\tabcolsep}{3.5pt}
\begin{tabular}{ccccc}
\hline
$a$ 
& $\lambda$ 
& $r_{\mathrm{pore}}$ 
& $\bar{\omega}_{\mathrm{eff}}$ 
& $P(90^\circ)\,\bar{\omega}_{\mathrm{eff}}$ \\
$[\mu\mathrm{m}]$ 
& [mm] 
& $[\mu\mathrm{m}]$ 
& range 
& range \\
\hline
100  & 0.87 & 5--18  & 0.856--0.862 & 0.851--0.858 \\
100  & 1.0  & 5--18  & 0.840--0.845 & 0.837--0.843 \\
100  & 3.0  & 5--18  & 0.439--0.445 & 0.439--0.445 \\
100  & 7.0  & 5--18  & 0.113--0.116 & 0.113--0.116 \\
1000 & 0.87 & 20--90 & 0.665--0.704 & 0.032--0.327 \\
1000 & 1.0  & 20--90 & 0.744--0.762 & 0.110--0.243 \\
1000 & 3.0  & 20--90 & 0.941--0.944 & 0.378--0.417 \\
1000 & 7.0  & 20--90 & 0.9746--0.9753 & 0.953--0.960 \\
\hline
\end{tabular}
\end{table}

To assess how the $D_f=2.1$ FA results depend on the adopted monomer size, we carried out additional \texttt{MSTM} calculations for the $D_f=2.1$ models at $a=10$ and $100\,\mu\mathrm{m}$, for $\lambda=0.87$, $1.0$, $3.0$, and $7.0\,\mathrm{mm}$. For each $(a,\lambda)$ pair, we kept the aggregate volume fixed and varied the monomer radius by changing $N_{\mathrm{mon}}$ from 8 to 8192, corresponding to $a_{\mathrm{mon}}=5.0$--$0.50\,\mu\mathrm{m}$ for $a=10\,\mu\mathrm{m}$ and $a_{\mathrm{mon}}=50$--$5.0\,\mu\mathrm{m}$ for $a=100\,\mu\mathrm{m}$. These sizes were chosen to make the full \texttt{MSTM} calculations computationally feasible while still probing the direction and magnitude of the monomer-size effect. Table~\ref{tab:appendixC_monomer_size} summarizes the ranges obtained across the explored monomer radii. For $a=10\,\mu\mathrm{m}$, the absolute scattering signal remains small across the explored monomer sizes. For $a=100\,\mu\mathrm{m}$, reducing $a_{\mathrm{mon}}$ from $50$ to $5.0\,\mu\mathrm{m}$ lowers $\bar{\omega}_{\mathrm{eff}}$ substantially, especially at $0.87$--$1.0\,\mathrm{mm}$, while $P(90^\circ)$ itself changes only weakly. Since $P(90^\circ)\simeq1$ in these tests, the behavior of $P(90^\circ)\,\bar{\omega}_{\mathrm{eff}}$ is controlled mainly by $\bar{\omega}_{\mathrm{eff}}$. These full-\texttt{MSTM} tests do not reach submicron monomers for the larger aggregates, which would require more than 8192 monomers; therefore, the high-$a$ \texttt{MSTM} aggregate results should be interpreted as dependent on the adopted monomer size.

\begin{table}[ht]
\centering
\caption{Monomer-size test for the $D_f=2.1$ FA models.}
\label{tab:appendixC_monomer_size}
\setlength{\tabcolsep}{3.5pt}
\begin{tabular}{ccccc}
\hline
$a$ 
& $\lambda$ 
& $a_{\mathrm{mon}}$ 
& $\bar{\omega}_{\mathrm{eff}}$ 
& $P(90^\circ)$ \\
$[\mu\mathrm{m}]$ 
& [mm] 
& $[\mu\mathrm{m}]$ 
& range 
& range \\
\hline
10  & 0.87 & 0.5--5.0  & $(1.31$--$1.46)\times10^{-2}$ & 0.991--1.000 \\
10  & 1.0  & 0.5--5.0  & $(9.73$--$10.8)\times10^{-3}$ & 0.991--1.000 \\
10  & 3.0  & 0.5--5.0  & $(0.86$--$1.66)\times10^{-3}$ & 0.991--1.000 \\
10  & 7.0  & 0.5--5.0  & $(0.085$--$1.20)\times10^{-3}$ & 0.991--1.000 \\
100 & 0.87 & 5.0--50  & 0.458--0.867 & 0.971--0.999 \\
100 & 1.0  & 5.0--50  & 0.455--0.849 & 0.977--0.999 \\
100 & 3.0  & 5.0--50  & 0.354--0.469 & 0.990--1.000 \\
100 & 7.0  & 5.0--50  & 0.113--0.127 & 0.990--1.000 \\
\hline
\end{tabular}
\end{table}
\bibliographystyle{aasjournalv7}
\bibliography{sample701}

\end{document}